# Datacubes of Hα, CaII K, CaII H and Hε line profiles of the full solar disk recorded daily at Meudon observatory since 2017 and some typical profiles of solar features

*Jean-Marie Malherbe*

Emeritus astronomer, Observatoire de Paris, PSL Research University, LIRA, France
Email: Jean-Marie.Malherbe@obspm.fr; ORCID: https://orcid.org/0000-0002-4180-3729

12 May 2026

## ABSTRACT

Systematic observations of the Sun are performed at Meudon observatory since 1908 under the form of monochromatic images. However, a major technical improvement occurred in 2017; since this date, spectroscopic datacubes are obtained daily with the spectroheliograph and a fast CCD camera. Line profiles of Hα (6562.8 A), CaII K (3933.7 A), CaII H (3968.5 A) and Hε (3970.1 A in the wing of CaII H) are recorded over the full solar disk under the form of 3D FITS files (x, y, λ). The optical spectral resolution is 0.15 A for Calcium and Hε (0.093 A/pixel), and 0.25 A for Hα (0.155 A/pixel); the spatial sampling is about 1 arc sec (the usual seeing is 2 arcsec). Datacubes are freely available since July 2017 in raw TIF (level 0) or processed FITS (level 1) format. Access to observations and typical line profiles associated to solar features are presented.

## I – INTRODUCTION

Systematic observations of the Sun started at Meudon in 1908 with Deslandres's spectroheliograph (Malherbe, 2024a). D'Azambuja was in charge of organizing the service and in parallel developed scientific research (Malherbe, 2023, 2024b). The spectroheliograph is a low dispersion grating instrument providing 0.15 A spectral resolution in order 5 (for lines in the violet part of the spectrum) and 0.25 A in order 3 (in the red part). Observations were made until 2001 with a thin slit in the spectrum selecting the line of interest; the detector was a photographic plate, it translated in the focal plane in perfect synchronism with the 25 cm scanning objective. A few monochromatic images, such as CaII K3 and Hα for the chromosphere, or CaII K1v for the photosphere, were produced daily (figure 1). The first CCD camera (100 x 1340 at 1 MHz) was incorporated in 2001 and the selecting slit in the spectrum was removed. As the readout speed was slow, we recorded only 5 spectral points in line profiles of Hα and CaII K. In 2017, this CCD was replaced by a fast and low noise sCMOS sensor (2048 x 2048 at 100 MHz) allowing to record hundreds of spectral points in line profiles (Malherbe & Dalmasse, 2019). We finally chose a compromise between the spectral FOV, the throughput of the camera and the volume of the data. CaII K, CaII H and Hε (in the red wing of CaII H) are observed simultaneously and we keep 100 spectral points in the line profiles (2048 x 2048 x 100 datacubes); 40 points are available for Hα (2048 x 2048 x 40 datacubes). Exposure times vary from 10 to 60 ms, so that the scan duration lasts between 20 seconds to 2 minutes. The spectral sample is 0.093 A/pixel at order 5, while it is 0.155 A/pixel at order 3. Hence, this is a low spectral resolution system (in comparison to large solar spectrographs and telescopes providing 10 times better spectral sampling), but as a counterpart, the spectroheliograph sees the full solar disk and prominences around. The spatial sampling is 1 arcsec/pixel, this is well adapted to the usual Meudon seeing of 2 arcsec. Long exposure observations are done for prominences with a neutral density 1 attenuator filtering the solar disk, but not the limb.

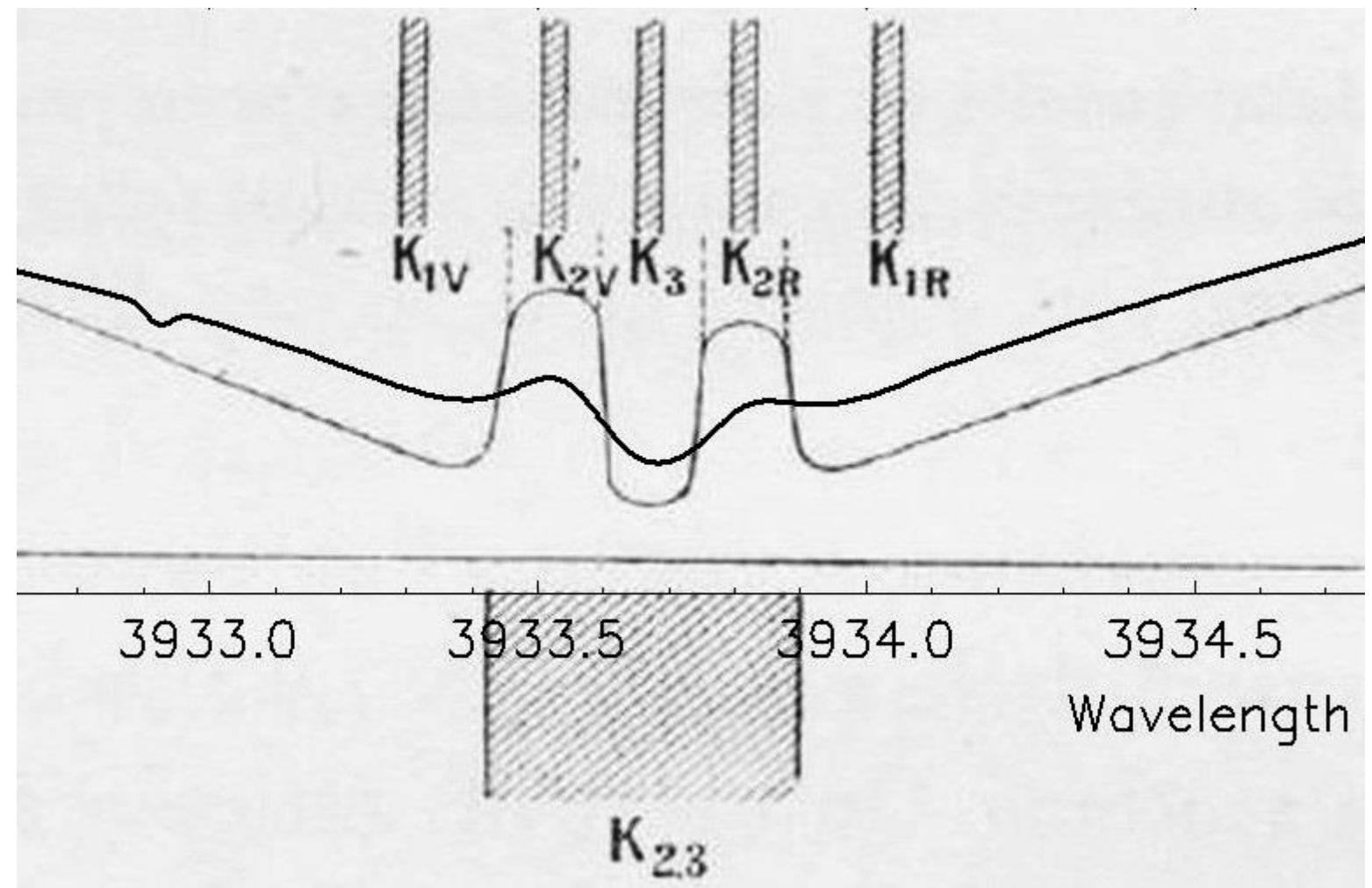

*Figure 1: the CaII K line profile as described by Deslandres. K3 is the line centre; K2v and K2r are adjacent peaks (± 0.2 A apart, v for violet wing, r for red wing); K1v and K1r are far wings. The spectroheliograph slit in the spectrum was able to select these components; however, the first version of 1908 integrated together K2v, K3 and K2r which was called K23.*

The spectroheliograph provides raw (level 0) datacubes in 16 bit integer TIF format together with dark currents that are stored on line. These are the reference files from which processed FITS datacubes (level 1) are derived for scientific purpose. Level 1 FITS data are corrected from dark current, line inclination and line curvature assuming a parabolic adjustment; they are also corrected of the coelostat rotation and the P angle rotation in order to have permanently the solar north up (in the y direction). These corrections have the advantage to deliver datacubes that can be directly used for scientific use; but the disadvantage is that corrections are not perfect and involve many interpolations, so that the final spectral and spatial resolutions are a little bit degraded in comparison with raw data. The FITS datacubes are (x, y, λ) organized, this is a set of monochromatic slices or scans at constant λ. On the contrary, raw data are (λ, y, x) organized, but the spectral line in slices (λ, y) is not parallel to the y axis and a little bit curved, so that the raw datacubes, when (x, y, λ) re-organized, are not a set of monochromatic slices. We keep online raw (level 0) TIF datacubes in order to provide a way to researchers willing to implement more sophisticated corrections than those systematically used to produce the standard FITS (level 1) datacubes.

CaII K, CaII H & Hε and Hα datacubes (TIF level 0 and FITS level 1) are systematically produced daily since July 2017 (of course weather permitted) and available on line.

## II - DESCRIPTION OF THE DATACUBES (RAW/TIF AND PROCESSED/FITS DATA)

Raw/TIF (level 0) data (λ, y, x) are not corrected from anything, such as dark current, line curvature and inclination, or solar P and coelostat angles to have the north up. For that reason, the precision of the line profiles is the best as there is no interpolation. They are available in yearly/monthly catalogues, in 3D 16 bit unsigned integer TIF format, here:

https://bass2000.obspm.fr/bass2000/data/pub/meudon/spc/tiff/

The filenames contain the date and UT time of observations. Most TIF files are compressed.

H* files: CaII H and Hε line profiles (100 wavelengths x 2048 x 2048 spatial)

K* files: CaII K line profiles (100 wavelengths x 2048 x 2048 spatial)

HA* files: Hα line profiles (40 wavelengths x 2048 x 2048 spatial)

Special observations are done for limb prominences (P) with a ND1 neutral density above the disk:

HP* files: long exposure CaII H and Hε line profiles (100 wavelengths x 2048 x 2048 spatial)

KP* files: long exposure CaII K line profiles (100 wavelengths x 2048 x 2048 spatial)

HAP* files: long exposure Hα line profiles (40 wavelengths x 2048 x 2048 spatial)

All D* files are corresponding dark currents (2D files, DHA*, DH*, DK*, DHAP*, DHP*, DKP*).

Each TIF file (λ, y, x) is accompanied by a TXT file containing observing parameters.

Processed/FITS (level 1) data are corrected from dark current, line curvature and inclination, and also from P angle and coelostat C angle in order to have the north up; this is the standard procedure delivering level 1 data, under the form of 3D (x, y, λ) 16 bit unsigned integer FITS format. As many interpolations are done by the processing code, the photometry of line profiles is a little bit degraded in comparison to raw/TIF level 0 datacubes, but FITS files are much more convenient for straightforward scientific use. FITS datacubes are available in yearly/monthly catalogues here:

For disk Hα, CaII H & Hε and CaII K:

https://bass2000.obspm.fr/bass2000/data/pub/meudon/spc/Ha/

https://bass2000.obspm.fr/bass2000/data/pub/meudon/spc/H/

https://bass2000.obspm.fr/bass2000/data/pub/meudon/spc/K/

For prominences (p), long exposure Hα, CaII H & Hε and CaII K (with ND1 disk attenuator):

https://bass2000.obspm.fr/bass2000/data/pub/meudon/spc/Hap/

https://bass2000.obspm.fr/bass2000/data/pub/meudon/spc/Hp/

https://bass2000.obspm.fr/bass2000/data/pub/meudon/spc/Kp/

The 3D FITS level 1 datacubes are compressed; filenames contain the strings “3D” and “fullprofile”. The number of spectral pixels (third dimension) is reduced in comparison with raw/TIF level 0 files, because of line inclination and curvature corrections (about 10% spectral pixels are lost). The Sun is rotated (bicubic interpolation).

## III - EXAMPLES OF SPECTRAL IMAGES DERIVED FROM THE DATACUBES

Below (figure 2) are provided some examples of spectral images that are derived from the corrected FITS and inserted daily in the bass2000 database (https://bass2000.obspm.fr/home.php). Spectral images are slices of the 3D datacubes at constant wavelength that correspond to a specific altitude in the photosphere or in the chromosphere. In particular, line wings form deeper in the solar atmosphere (photosphere or low chromosphere) than line cores (chromosphere). For that reason, filaments visible in line cores are no more visible in the wings, except in the case of strong radial velocities which shift the line towards longer or shorter wavelengths.

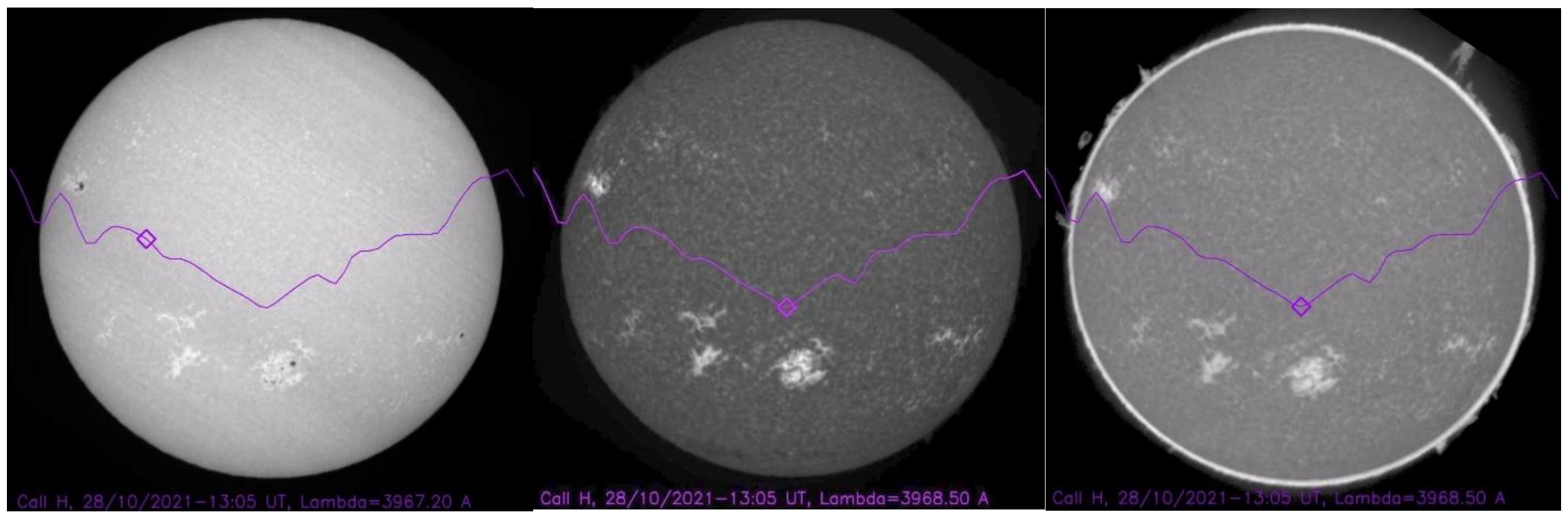

*CaII H1v (left), CaII H3 (centre), CaII H3 long exposure for prominences (right) with neutral density ND1 upon the disk, 28 October 2021*

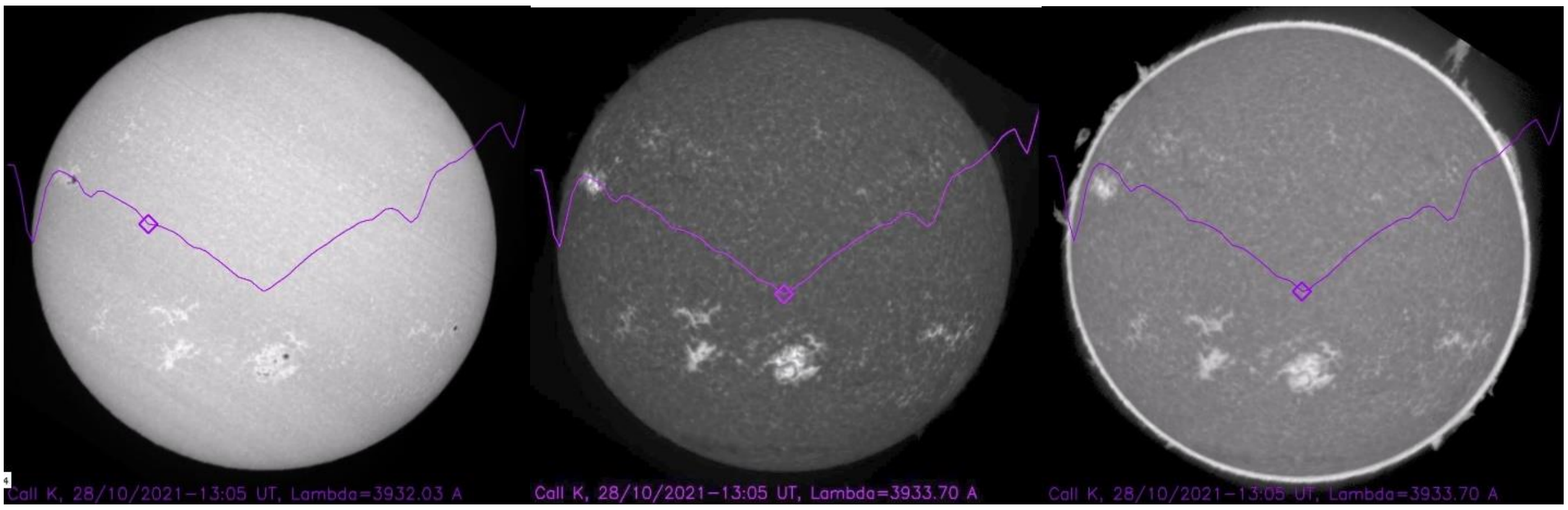

*CaII K1v (left), CaII K3 (centre), CaII K3 long exposure for prominences with ND1 (right)*

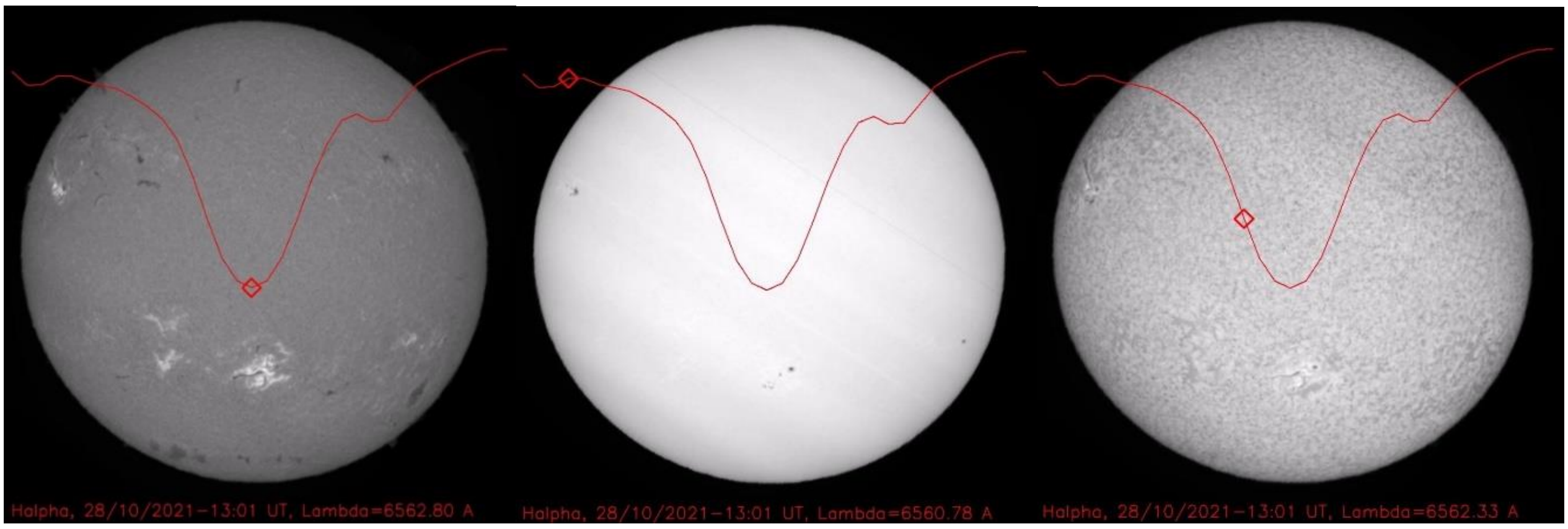

*Figure 2: Hα centre (left), H blue continuum (centre), H blue wing (right), 28 October 2021*

## IV – TYPICAL PROFILES OF SOLAR FEATURES DERIVED FROM THE DATACUBES

The datacubes, systematically recorded since almost 10 years, are a useful tool to explore the spectral line behaviour in various structures of the Sun, such as active regions, filaments, sunspots, faculae or prominences at the limb. Cretignier *et al* (2023) intensively used Meudon CaII H & K datacubes and compared them to higher spectral resolution observations. More recently, Tamburri *et al* (2026) reported the first high spectral and spatial resolution observations of CaII H & Hε with the giant DKIST telescope. We first present CaII H and Hε observations got with Meudon spectroheliograph (figures 3 & 4) on 10 May 2024 prior to a series of major flares in AR 3664 (X1 on 9 May at 17:30 UT, X4 on 10 May at 06:30 UT, X9 on 14 May at 16:45 UT). The 10 May flare is at the origin of the large solar disturbances and aurorae visible in many places and at low latitudes on 11 May. Figure 3 shows monochromatic images of AR 3664 which produced the flares, revealing many bright points or flare kernels both in CaII H3 and Hε. Figure 4 displays line profiles of two bright points with emission in CaII H3, CaII K3 and Hε.

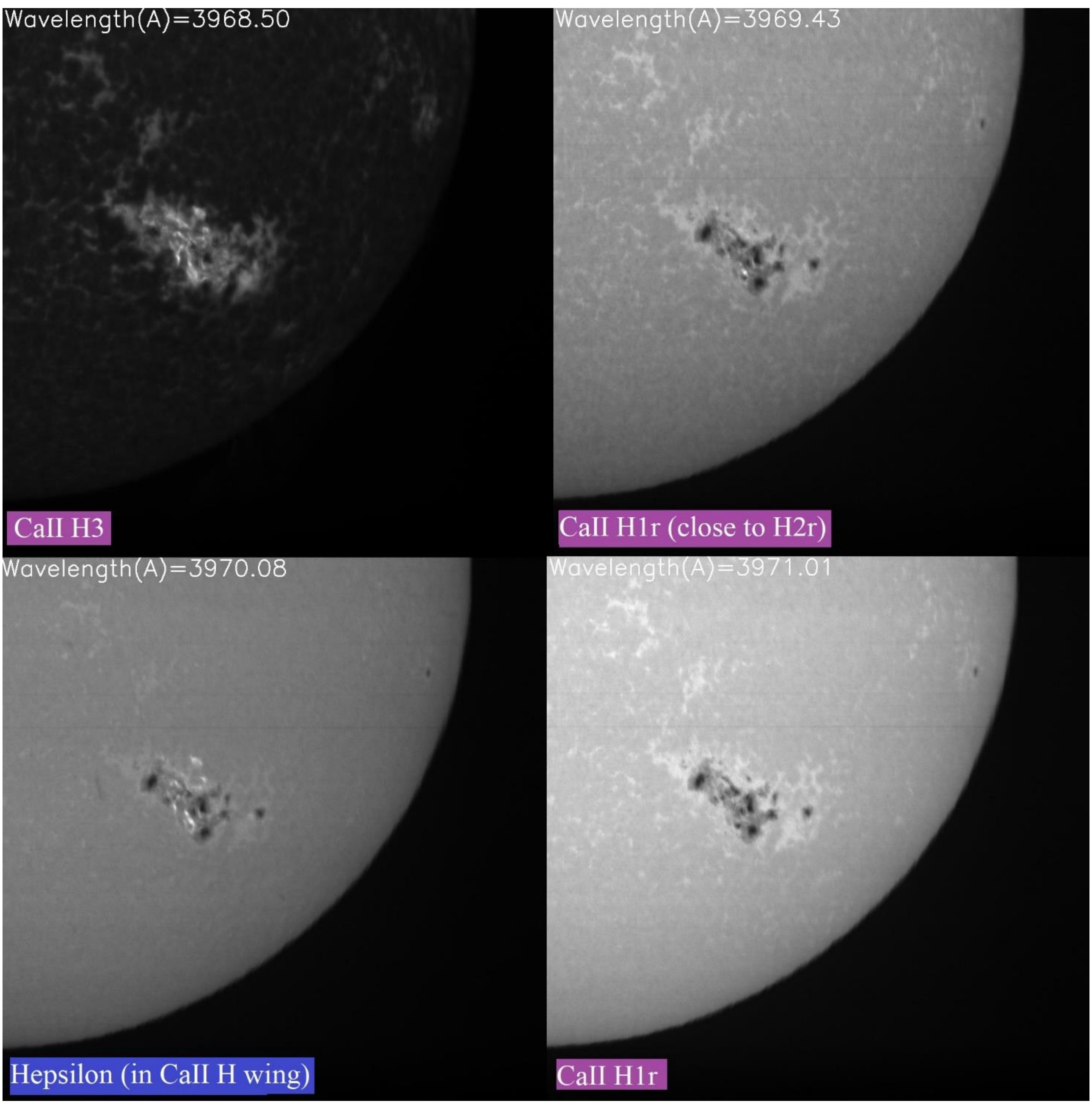

*Figure 3 : 10 May 2024, 06 :25 UT, AR 3664, CaII H3, CaII K1r (close do H2r), Hε and CaII H1r.*

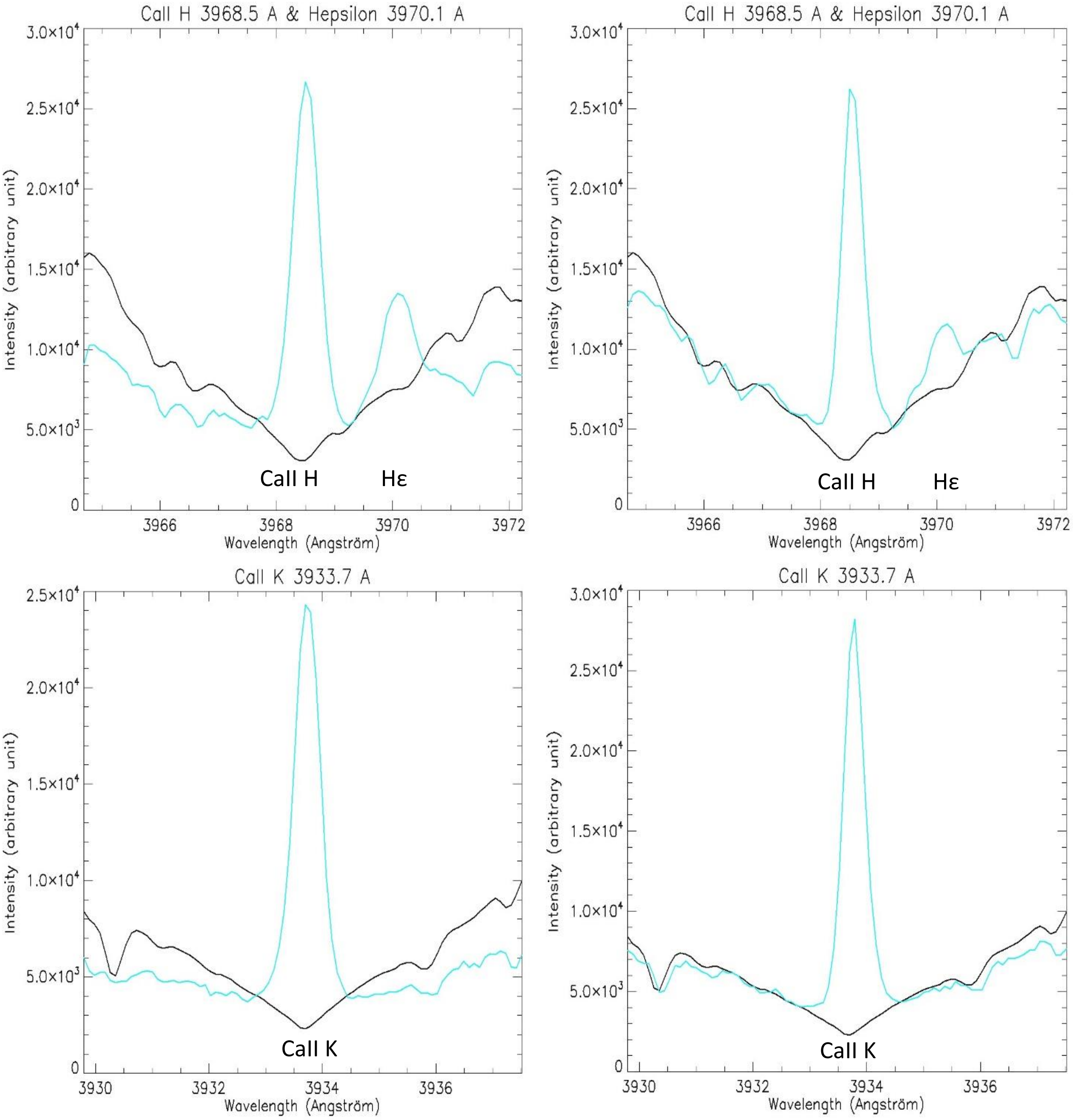

*Figure 4 : line profiles of 10 May 2024, 06 :25 UT, CaII H (top) and CaII K (bottom) line profiles of two bright points of AR 3664 (reported as points 1 & 2 in Figure 5). Black line: profiles at disk centre. Blue line: bright point line profiles.*

Figure 4 shows the line profiles of two **bright points** of AR 3664 (points 1 & 2 reported in Figure 5 below) with intense emission in the core of CaII H and CaII K lines; it reveals also emission in Hε, which is in the quiet Sun a very faint line in absorption in the red wing of CaII H. This phenomenon has been observed with the CHROMIS instrument at the Swedish SST telescope by Rouppe van der Voort *et al* (2024) and modelled by Krikova *et al* (2023). They emphasize the importance of Hε emission as the signature of heating processes of the solar atmosphere in relation to Ellerman bombs and magnetic reconnection phenomena at small spatial scales. Our results of Figure 4 are very close to the ones presented by Tamburri *et al* (2026) in their own figure 5 with DKIST/VISP in a C class flaring region.

Figure 5 below shows line profiles observed in **sunspots** of AR 3664 with Meudon spectroheliograph, indicated by points 4 & 6, with small emission in CaII H3 and K3.

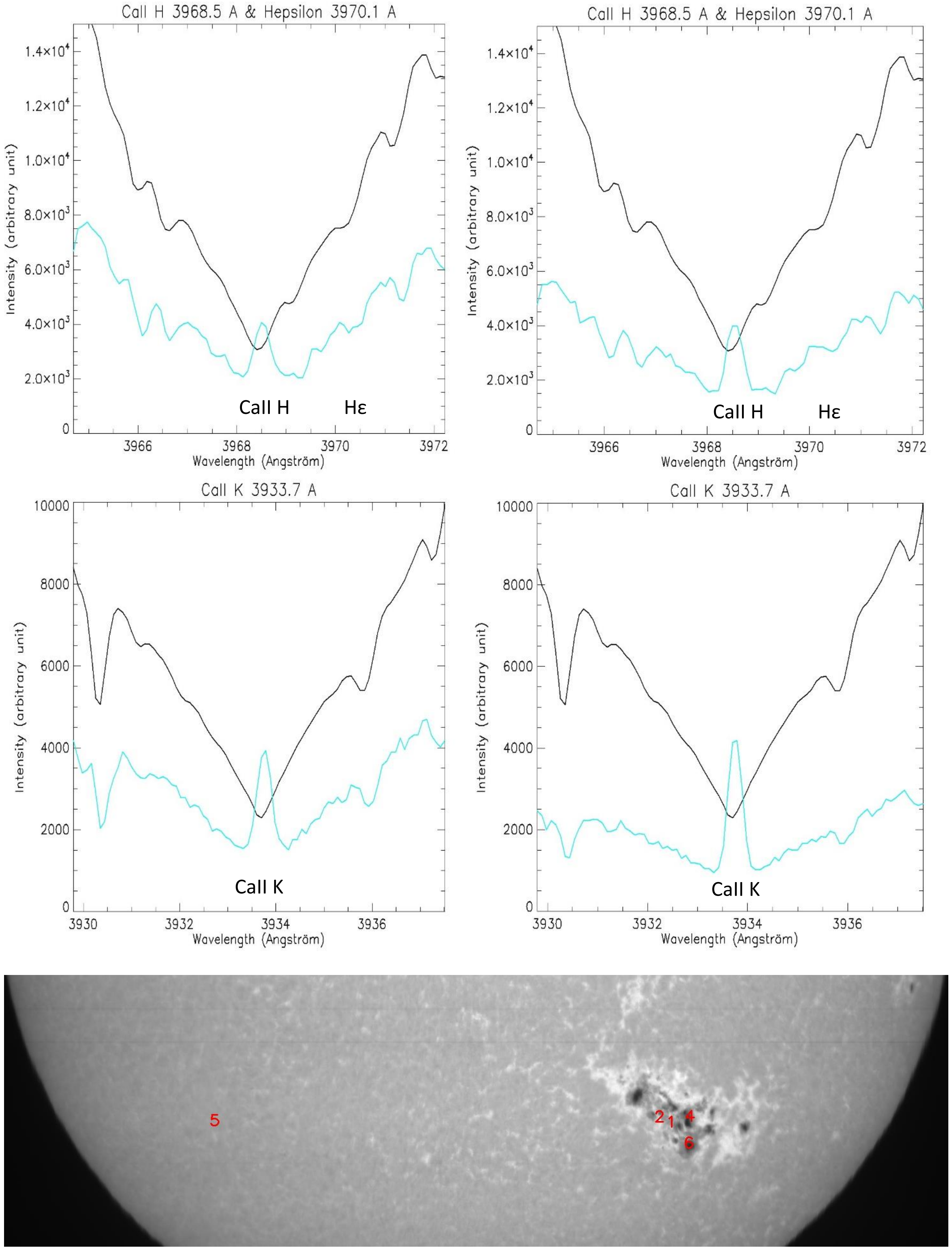

*Figure 5 : line profiles of 10 May 2024, 06 :25 UT, CaII H (top) and CaII K (middle) line profiles of two sunspots (points 4 & 6, bottom) of AR 3664. Black line: profiles at disk centre. Blue line: dark sunspot umbra line profiles. Hε remains in absorption.*

Figure 6 below shows line profiles observed in a bright **facular** region (point 3) and in a **filament** (point 5) with the spectroheliograph, with small facular emission in CaII H3 and K3. Hε is in emission in the facula, and in absorption in the filament. K2 and H2 peaks appear in the facula.

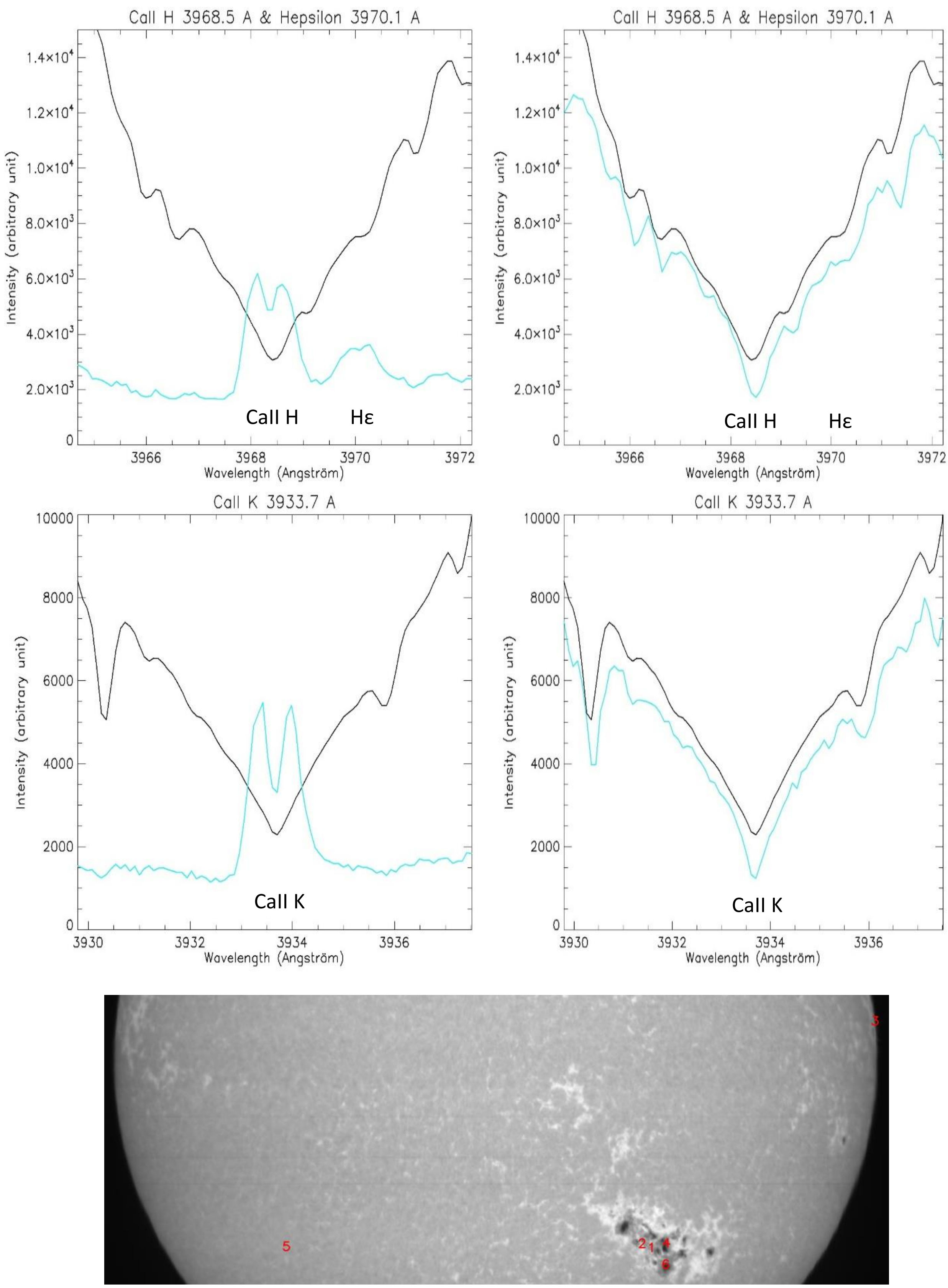

*Figure 6 : line profiles of 10 May 2024, 06 :25 UT, CaII H (top) and K (middle) profiles of points 3 (left) & 5 (right). Black line: profiles at disk centre. Blue line: facula or filament profiles.*

Figure 7 shows other line profiles observed in **bright points** on 3 October 2024 (points 1 & 2) with the spectroheliograph, again in emission in CaII H3 and in Hε.

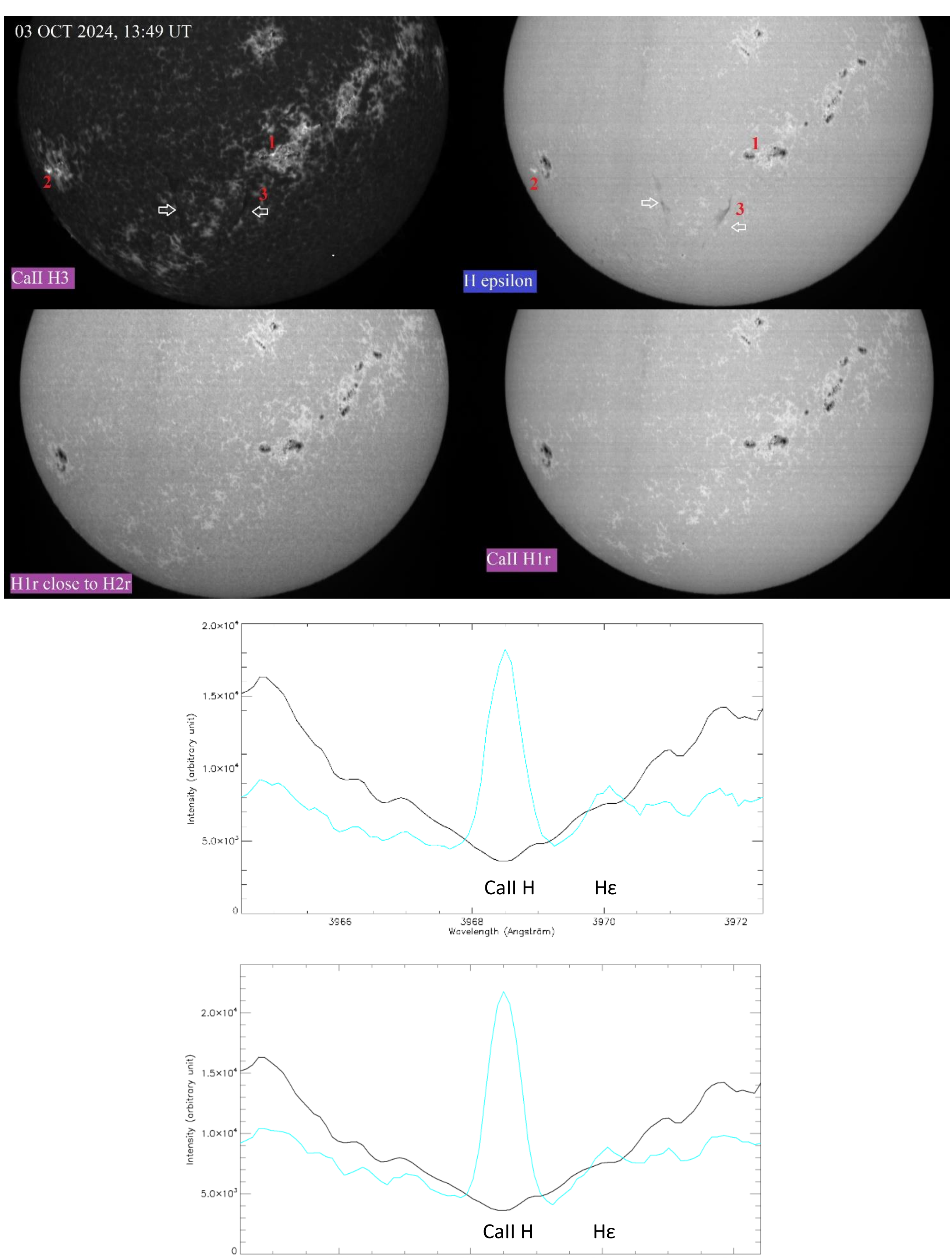

*Figure 7 : line profiles of 3 October 2024, 13:49 UT, CaII H line profiles of points 1 (top) & 2 (bottom). Black line: profiles at disk centre. Blue line: bright point profiles.*

Figure 8 displays other line profiles observed in **filaments** on 3 October 2024 (points 3 and arrows of Figure 6) and the northern filament of 10 October 2024 with the spectroheliograph. It is visible in absorption both in CaII H3 and in Hε. However, only densest parts appear in Hε.

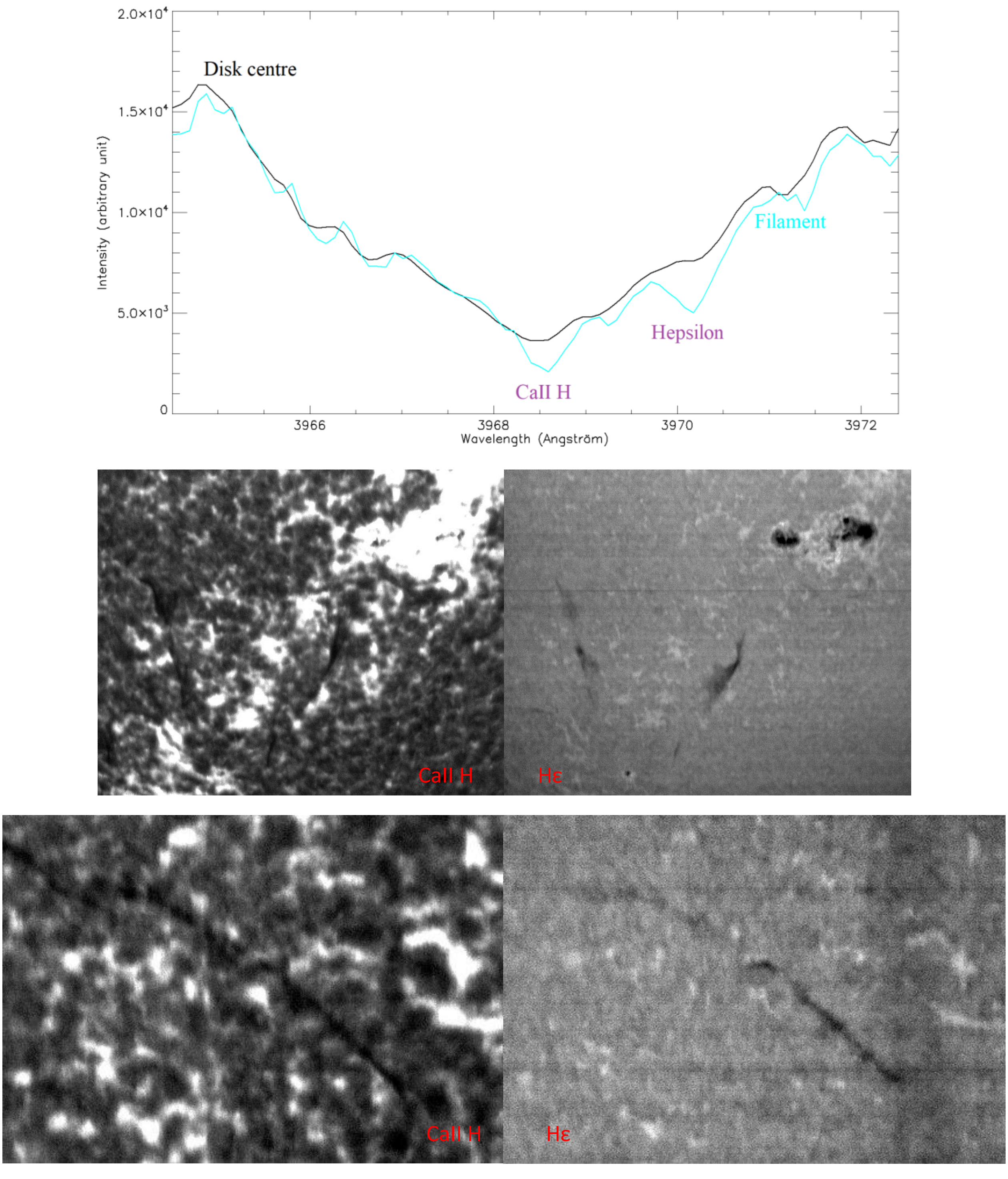

*Figure 8 : filament of 3 October 2024, 13:49 UT, CaII H profiles (top) and images (CaII H3 and Hε, middle). Black line: profiles at disk centre. Blue line: filament. Bottom: images only (CaII H3 at left, and Hε at right) of the northern filament of 10 October 2024.*

We also looked at data of 6 November 2017, prior to a large solar flare (X17) and focused on **bright points** in the active region of Figure 9, and this corroborates what we saw with other events. There is a strong emission in CaII H3 together with emission of Hε which appear both in monochromatic images and spectra. For comparison, the Hα line is not so reactive (Figure 10).

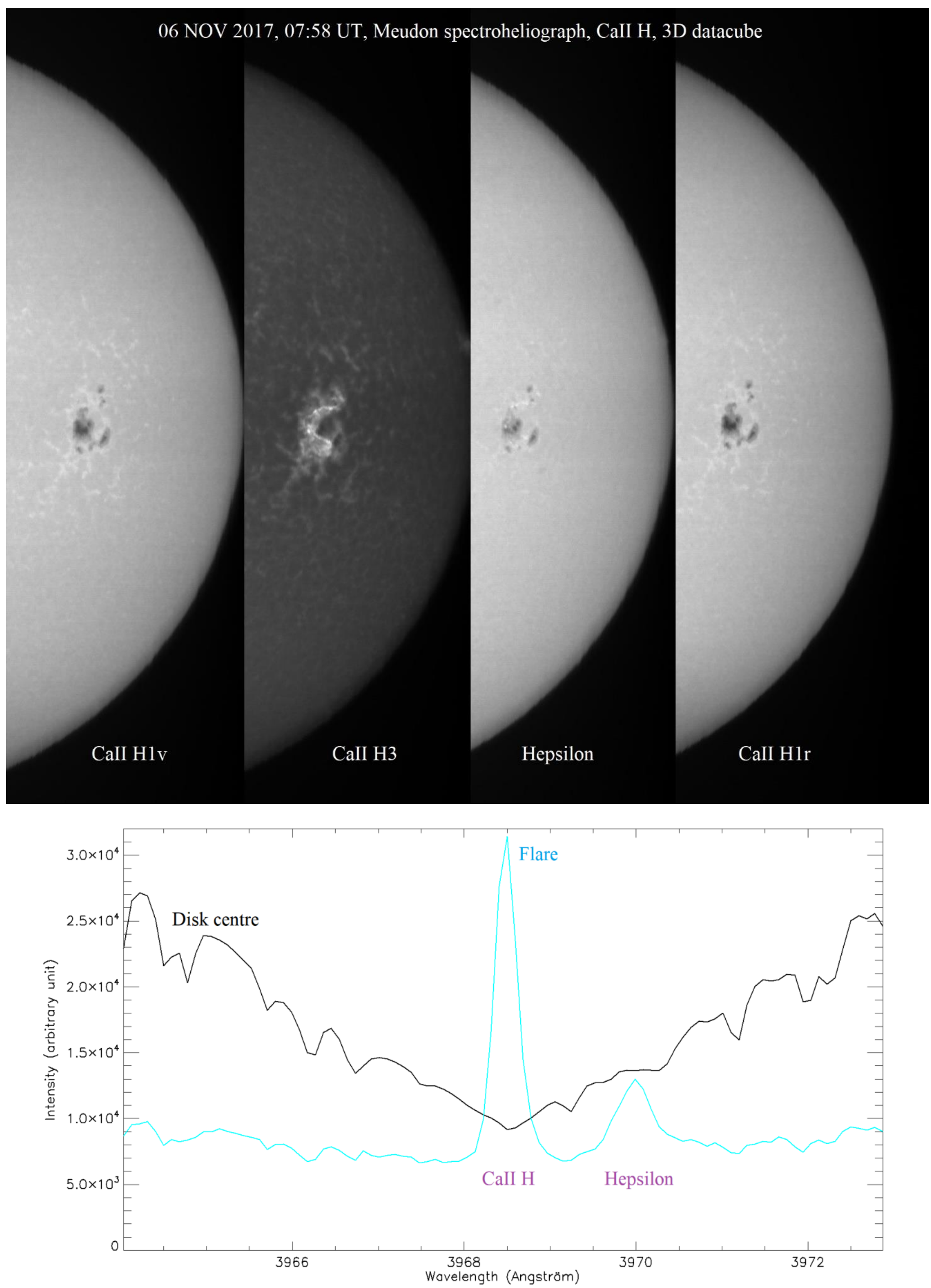

*Figure 9 : flaring active region of 6 November 2017, 07:58 UT, CaII H1v, CaII H3, Hε and CaII H1r images (top) together with CaII H line profiles of a typical bright point. Black line: profiles at disk centre. Blue line: bright point profile.*

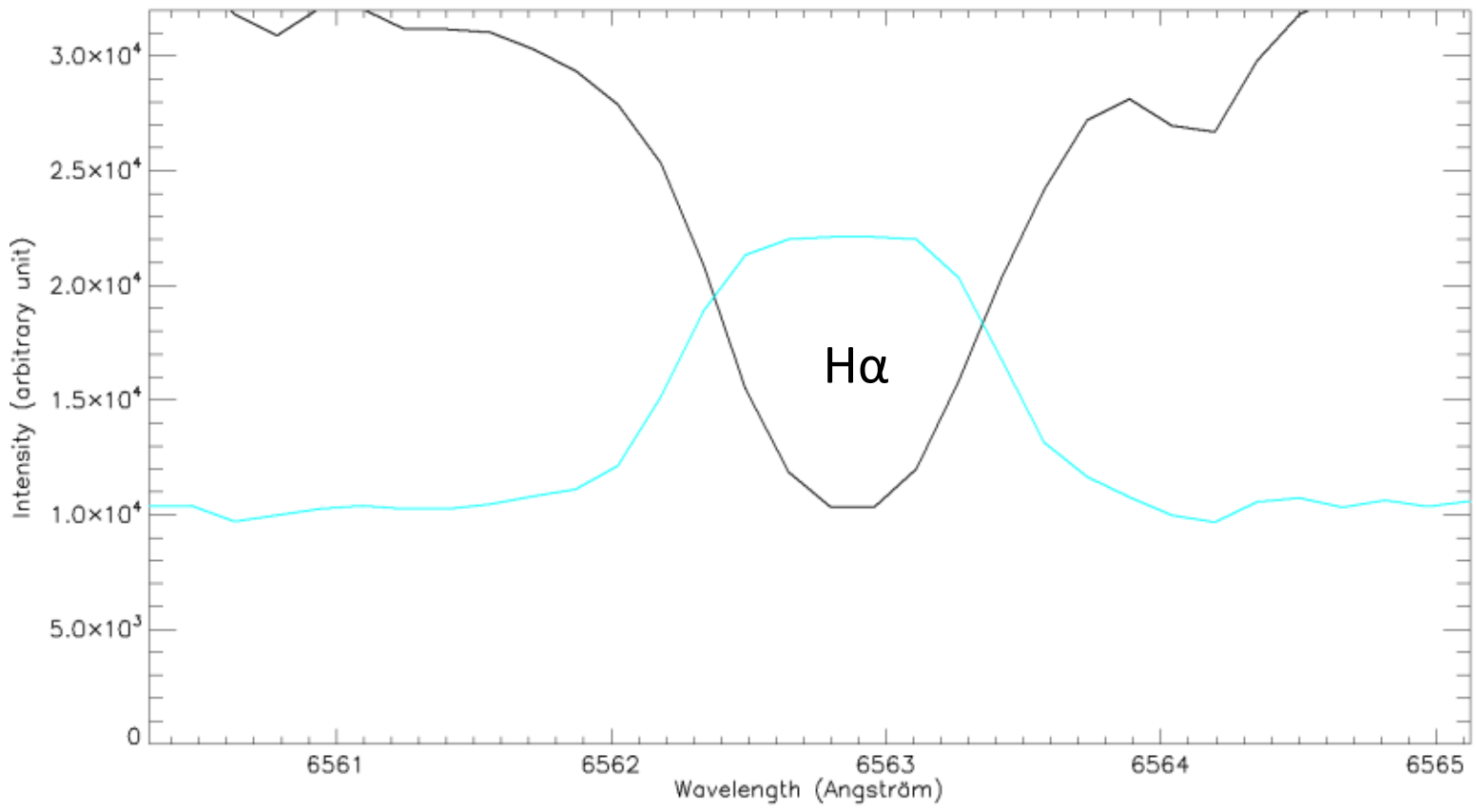

*Figure 10: flaring active region of 6 November 2017, 07:58 UT, Hα line profile of a typical bright point. Black line: profiles at disk centre. Blue line: bright point profile.*

At last, we also looked at **prominences** of 10 May 2024 (Figure 11). CaII H3 together with Hε are in emission, but Hε is very faint and only densest parts of prominences are visible.

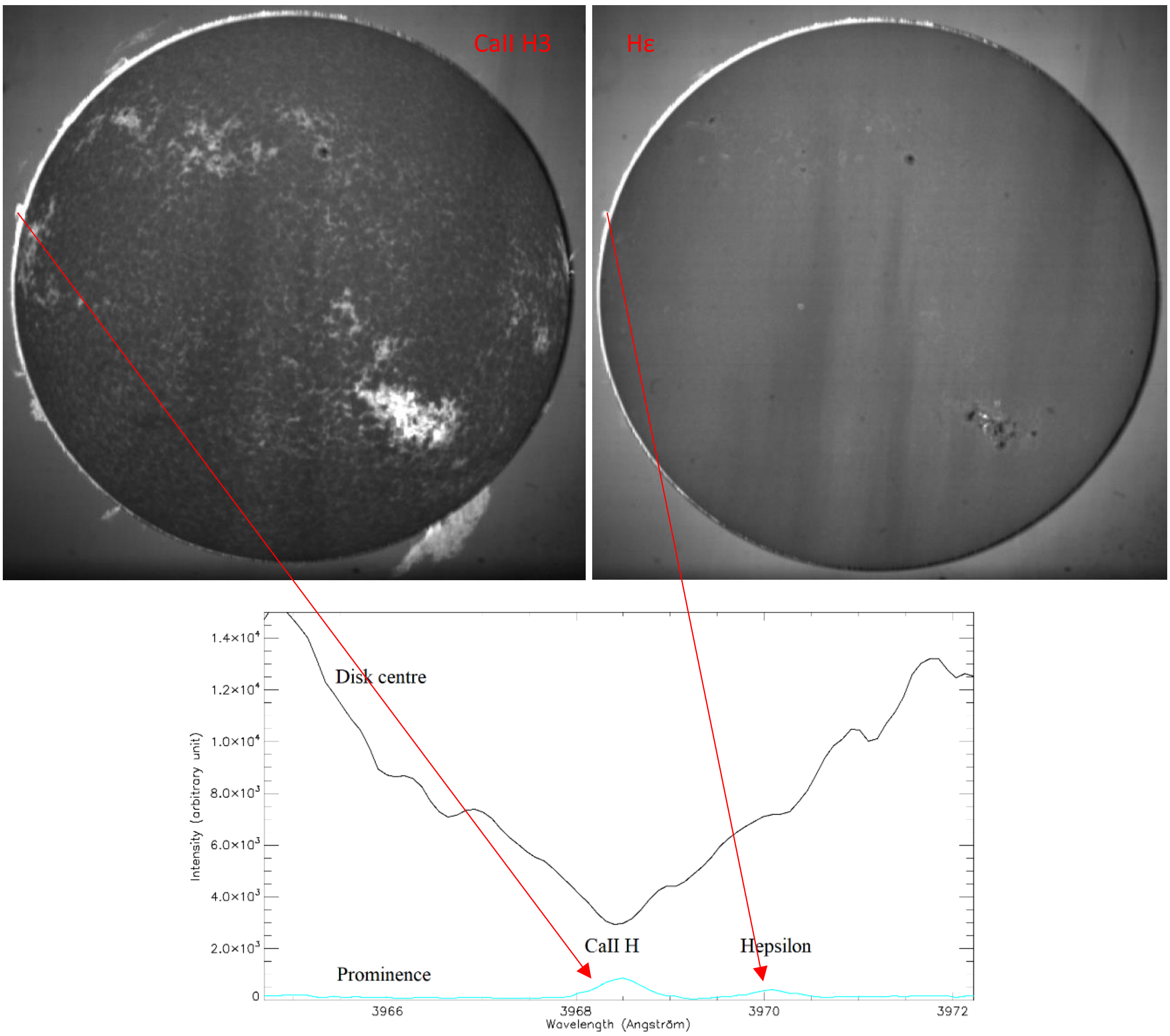

*Figure 11: prominences of 10 May 2024. CaII H3 and Hε images (top). Typical line profile showing emission of CaII H and Hε (right) in a dense prominence. There was also a huge prominence south west of the Sun in CaII H, but it has no signature in Hε.*

## V – CONCLUSION

Hα (6562.8 A), CaII K (3933.7 A), CaII H (3968.5 A) and Hε (3970.1 A in the wing of CaII H) line profiles over the full solar disk are recorded daily since July 2017 with Meudon spectroheliograph under the form of processed 3D FITS level 1 files (x, y, λ) for immediate scientific use, and raw TIF level 0 files for specific data treatments. Datacubes are available freely on line to the solar community. The spectral resolution is moderate (0.093 A/pixel in the violet, 0.155 A/pixel in the red) but the entire disk is available with 1 arcsec pixel, so that all solar features are available at the same time. The usual cadence is a few datacubes per day; however, the scans are fast (20 s to 2 min) so that the cadence can be easily increased upon request to the observers, for scientific programs needing temporal resolution, as in the case of forecasted solar activity.

## VI – REFERENCES

Cretignier, M., Pietrow, A. G. M., Aigrain, S., 2023, "Stellar surface information from the CaII K & K lines. I. Intensity profiles of the solar activity components", *arXiv*: 2310.15926v1

Krikova, K., Pereira, T.M.D., Rouppe van der Voort, L.H.M., 2023, "Formation of Hε in the solar atmosphere", *Astron. Astrophys*., 677, A52

Malherbe, J.M., Dalmasse, K., 2019, "The new 2018 version of Meudon Spectroheliograph", *Solar Physics*, 294, 52

Malherbe, J.M., 2023, "Optical characteristics and capabilities of the successive versions of Meudon spectroheliograph (1908-2023)", *arXiv*:2303:10952v1

Malherbe, J.M., 2024a, "The quadruple spectroheliograph of Meudon observatory (1909-1959)", *arXiv*:2402:16436v1

Malherbe, J.M., 2024b, "Lucien and Marguerite D'Azambuja, Explorers of Solar Activity (1899–1959), *JAHH*, 27, 303

Rouppe van der Voort, L.H.M., Joshi, J., Krikova, K., 2024, "Observations of magnetic reconnection in the deep solar atmosphere in the Hε line", *arXiv*:2401.12077v1

Tamburi, C., Kowalski, A., Cauzzi, G., Kazachenko, M., Tritschler, A., Yadav, R., French, R., Notsu, Y., Reardon, K., Tristan, I., 2026, "Spectroscopic analysis and modelling of the first CaII H & Hε flare spectra from DKIST/VISP", *arXiv*:2602.15658v1